\documentclass[aps,prd,twocolumn,showpacs,showkeys,preprintnumbers,amsmath,amssymb]{revtex4-1}
\usepackage{amssymb}
\usepackage{amsfonts}
\usepackage{amsmath}

\begin{document}

\title{The Noncommutative Harmonic Oscillator based in Symplectic Representation of Galilei Group}
\author{R.G.G. Amorim}
\email{ronniamorim@gmail.com} \affiliation{Instituto Federal de Educa\c{c}\~{a}o, Ci\^{e}ncia e Tecnologia de Goi\'{a}s, Campus de Luzi\^{a}nia, 72811-580, Luzi\^{a}nia, GO, Brazil.}

\author{S. C. Ulhoa }
\email{sc.ulhoa@gmail.com} \affiliation{Faculdade Gama,Universidade de Bras\'{i}lia, 72444-240, Setor Leste (Gama), Bras\'{i}lia-DF, Brazil.}

\author{A.E. Santana }
\email{asantana@unb.br} \affiliation{Instituto de F\'{\i}sica, Universidade de Bras\'{\i}lia, 70910-900, Bras\'{\i}lia, DF, Brazil.}

\begin{abstract}
We study symplectic unitary representations for the Galilei group and derive the Schr\"odinger
equation in phase space. Our formalism is based on the non-commutative structure of the star-product.
Guided by group theoretical concepts, we construct a physically consistent phase-space
theory in which each state is described by a quasi-probability amplitude associated with the Wigner
function. As applications, we derive the Wigner functions for the 3D harmonic oscillator and the
non-commutative oscillator in phase space.
\end{abstract}



\maketitle
\section{Introduction}

The concept of non-commutativity in Physics dates
back to the birth of Quantum Mechanics. Heisenberg's
uncertainty principle gave physical substance to that notion.
From a more mathematical viewpoint, the assumption
that spatial and momentum coordinates do not commute
has lead to non-commutative geometry, which has
offered new insight. As a matter of fact, Heisenberg himself
suggested that an uncertainty relation among the
spatial coordinates might avoid the singularities due to
particle self-energies~\cite{pauli1, pauli2, jackiw}. The first elaborate analysis in
this line of research is, however, due to Snyder~\cite{snyder1, snyder2}, a
former student of Oppenheimer, who proposed a new vision
of space-time. According to Snyder the space-time
should be understood as a collection of minimum-size
cells, forming a lattice structure rather than a continuum.
In this picture, it is inadequate to define spacetime
points, because non-commutativity bars accurate
measurements of particle positions.

Spatial non-commutativity can be introduced by means
of Hermitian operators standing for the space-time coordinates, $\widehat{x}^{\mu }$, and satisfying the algebra $[\widehat{x}^{\mu },\widehat{x}^{\nu }]=i\theta ^{\mu \nu }$, where the $\theta
^{\mu \nu }$ are the components of a constant antisymmetric
tensor. The commutation relations imply $\Delta \widehat{x}^{\mu }\Delta \widehat{x}%
^{\nu }\geq \frac{1}{2}|\theta ^{\mu \nu }|$, so that the non-commutativity becomes
relevant for distances of the order of $\sqrt{|\theta ^{\mu \nu }|}$. These
are the basics of non-commutative geometry.

Over the last decades, interest in this type of noncommutative
geometry progressively grew as applications were made to non-abelian theories ~\cite{chans}, gravitation~\cite{kalau, kastler, connes1}, the standard model~\cite{connes2, varilly1, varilly2}, and the Quantum Hall
effect~\cite{belissard}. The formalism is nonetheless still incomplete.
In particular, no Wigner-function analysis of non-commutativity
in phase space has been presented.

In 1932, in a development that was contemporary to
the initial studies of non-commutative geometry, Wigner
introduced a quantum formalism in the phase space $\Gamma $,
with a view to applications in quantum kinetic theory~\cite{wig1}.
His approach associates each operator $A$ in the
Hilbert space $\mathcal{H}$ with a function $a_{W}(q,p)$ defined in $\Gamma $~\cite%
{wig1, wig2, wig3, wig4}. The application
$\Omega _{W}:A\rightarrow
a_{W}(q,p)$ is such
that the associative algebra of operators in $\mathcal{H}$ defines an
associative non-commutative algebra in $\Gamma $.

The non-commutativity stems from the nature of
the product between two operators in $\mathcal{H}$. Given
two operators $A$ and $B$, we have the mapping $\Omega :AB\rightarrow
a_{W}(q,p)\star b_{W}(q,p)$. Here the (non-commutative)
star-product $\star $ is defined by the identity $a_{W}(q,p)\star
b_{W}(q,p)=a_{W}(q,p)\exp [\frac{i}{2}(\frac{\overleftarrow{\partial }}{%
\partial q}\frac{\overrightarrow{\partial }}{\partial p}-\frac{%
\overleftarrow{\partial }}{\partial p}\frac{\overrightarrow{\partial }}{%
\partial q})]b_{W}(q,p)$, the so-called
Moyal product.

Although the phase space and Moyal product have
been explored in different ways ~\cite%
{wig2, wig3, wig4, zak1, boo1, moy1, moy2, seb3, seb4, seb41, seb42, seb43,
seb5, seb8, seb9, seb10, seb11, seb12,seb13,bos,seb445,dodo6,dodo7,dodo8}, only recently
has a physically consistent representation theory been
developed. First, irreducible unitary representations of
kinematical groups in $\Gamma $ have been studied with operators
of type $\widehat{A}=a_{W}\star $ acting on the function $b_{W}$,
i. e.,$\widehat{A}b_{W}(q,p)=a_{W}(q,p)\star b_{W}(q,p)$. For the Galilei
group, this symplectic star-representation yields a phase-space
Schr\"odinger equation, the role of wave functions
(the quasi-amplitudes of probability) being played by
the Wigner functions (the quasi-distributions of probability)
~\cite{seb1}. This method affords a derivation of the Wigner
functions without the intricacies of the Liouville-von
Neumann equation, which provided the original starting point in Wigner's approach, and leads to a prescription
to derive symplectic star-representations for the Poincar\'e
symmetry. This in turn leads to phase-space representations
of the Klein-Gordon and Dirac fields~\cite{seb2, seb22}.
This algebraic formalism is a natural candidate to derive
Wigner functions for nonclassical radiation states and for
non-commutative space-time systems, such as the quantum
Hall effect.

Here we explore the elements of this approach. Considering
the Galilei group, we study the eigenvalue problem
of the phase-space Schr\"odinger equation to first
treat the three-dimensional harmonic oscillator and derive
the quasi-amplitude of probability and the corresponding
Wigner function. These results provide a starting
point for our analysis of nonclassical electromagnetic radiation
states and of phase-space Bose-Einstein condensation.
As a second application, we consider a 2-D
non-commutative harmonic oscillator, a prototype of the
Hall effect in phase space.

The presentation is organized as follows. In Section 2,
we define a Hilbert space $\mathcal{H}(\Gamma )$ over a phase space, including
a natural symplectic structure. We then take the
space $\mathcal{H}(\Gamma )$ as the carrier space for unitary representations
of the Galilei group. In Section 3, we construct the
generators $A_{w}(q,p)\star $ for the Galilei group, hence deriving
a representation for the phase-space Schr\"{o}dinger equation.
In Section 4, we present solutions of the Schr\"{o}dinger
equation in phase space for the three dimensional harmonic
oscillator. In Section 5, we consider the noncommutative
oscillator in phase space. In Section 6, we
present concluding remarks.

\section{Hilbert Space and Symplectic Structure}

Consider an analytical manifold $\mathbb{M}$, where each point is
specified by coordinates $q$. The coordinates of each point
in the cotangent-bundle $\Gamma =T^{\ast }\mathbb{M}$ are denoted $(q,p)$.
The 2N-dimensional manifold $\Gamma $ is equipped with a 2-
form, defined by $\omega
=dq\wedge dp$, called the symplectic form.
With the symplectic form, the operator

\begin{equation}
\Lambda =\frac{\overleftarrow{\partial }}{\partial q}\frac{\overrightarrow{%
\partial }}{\partial p}-\frac{\overleftarrow{\partial }}{\partial p}\frac{%
\overrightarrow{\partial }}{\partial q}\   \label{mar262}
\end{equation}%
leads to the Poisson bracket,
\begin{equation*}
\omega (f\Lambda ,g\Lambda )=\{f,g\}=f\Lambda g,
\end{equation*}%
where
\begin{equation*}
\{f,g\}=\frac{\partial f}{\partial q}\frac{\partial g}{\partial p}-\frac{%
\partial f}{\partial p}\frac{\partial g}{\partial q}\,.
\end{equation*}
Here $f=f(q,p)$ and $g=g(q,p)$.

The manifold $\Gamma =T^{\ast }\mathbb{M}$
endowed with this symplectic
structure is then called the phase space, and the algebraic set of the analytical functions $f(q,p)$ is denoted by $C^{\infty
}(\Gamma )$. The vector fields over $\Gamma $ are given by
\begin{equation*}
X_{f}=f\Lambda =\frac{\partial f}{\partial q}\frac{\partial }{\partial p}-%
\frac{\partial f}{\partial p}\frac{\partial }{\partial q}.
\end{equation*}

A Hilbert space associated with $\Gamma $ is introduced by a set of complex
functions, $\psi (q,p)$, which are square integrable in $C^{\infty }(\Gamma
) $, i.e.
\begin{equation*}
\int dpdq\psi ^{\dagger }(q,p)\psi (q,p)<\infty .\
\end{equation*}%
These functions may be then defined as $\psi (q,p)=\langle q,p|\psi \rangle $%
, with
\begin{equation*}
\int dpdq\ |q,p\rangle \langle q,p|=1,
\end{equation*}%
such that
\begin{equation*}
\langle \psi |\phi \rangle =\int dpdq\ \psi ^{\dagger }(q,p)\phi (q,p),
\end{equation*}%
where $\langle \psi |$ is a dual vector of $|\psi \rangle $. This Hilbert space, denoted
$\mathcal{H}(\Gamma ),$ is here the carrier space for representations
of Lie algebras.

Consider $\ell =\{a_{i},i=1,2,3,...\}$ a Lie algebra over the (real) field $%
\mathbb{R}$, of a Lie group $\mathcal{G}$, characterized by the algebraic
relations $(a_{i},a_{j})=C_{ijk}a_{k}$, where $C_{ijk}\in \mathbb{R}$ are
the structure constants and $(,)$ is the Lie product. We construct unitary
symplectic representations for $\ell ,$ denoted by $\ell _{Sp}$, using the
star-product. The associative product in $\mathcal{H}(\Gamma )$
is obtained from the operator $\Lambda $, in Eq.~(\ref{mar262}) as a mapping $e^{ia\Lambda
}=\star :$ \ $\Gamma \times \Gamma \rightarrow \Gamma ,$ defined the equality
\begin{eqnarray}
(f\star g)(q,p) &=&f(q,p)e^{ia\Lambda }g(q,p)  \notag \\
&=&\exp \left[ ia\left( \partial _{q}\partial _{p^{\prime }}-\partial
_{p}\partial _{q^{\prime }}\right) \right]  \notag \\
&&\times f(q,p)g(q^{\prime },p^{\prime })|_{q^{\prime },p^{\prime }=q,p},
\label{A9812}
\end{eqnarray}

where $f$ and $g$ are functions in $C^{\infty }(\Gamma )$ and $\partial
_{x}=\partial /\partial x$ $(x=p,q)$. The constant $a$, which fixes units, has no special
meaning. The usual associative product is obtained
by letting $a=0$.  To each function $f(q,p)$ an operator $\widehat{{f}}=f(q,p)\star $ is associated, which will be used as the generator
of unitary transformations.

\section{Galilei-Lie Algebra and Sch\"{o}dinger Equation in Phase Space}

We now study the representation of the Galilei group $\mathcal{H}%
(\Gamma )$, which leads us to the Schr\"odinger equation in phase
space, and connect this representation with the
Wigner formalism.

With the star-operator $\widehat{A}=a\star $, where $a=a(q,p)$, we
define a momentum- and a position-like operator by the
equalities
\begin{equation}
\widehat{Q}=q\star =q+\frac{i\hbar }{2}\partial _{p}\,,  \label{eq 7}
\end{equation}
and
\begin{equation}
\widehat{P}=p\star =p-\frac{i\hbar }{2}\partial _{q}\,,  \label{eq 8}
\end{equation}%
respectively.

We can then define a boost, an angular momentum,
and a Hamiltonian-like operator by the equalities
\begin{equation}
\widehat{K}=m\widehat{Q}_{i}-t\widehat{P}_{i}\,,  \label{eq 9}
\end{equation}%
\begin{eqnarray*}
\widehat{L}_{i} &=&\epsilon _{ijk}\widehat{Q}_{j}\widehat{P}_{k} \\
&=&\epsilon _{ijk}q_{j}p_{k}-\frac{i\hbar }{2}\epsilon _{ijk}q_{j}\frac{%
\partial }{\partial p_{k}} \\
&&+\frac{i\hbar }{2}\epsilon _{ijk}p_{k}\frac{\partial }{\partial q_{j}}+%
\frac{\hbar ^{2}}{4}\frac{\partial ^{2}}{\partial q_{j}\partial p_{k}}
\end{eqnarray*}
and%
\begin{eqnarray*}
\widehat{H} &=&\frac{\widehat{P}^{2}}{2m}=\frac{1}{2m}(\widehat{P}_{1}^{2}+%
\widehat{P}_{2}^{2}+\widehat{P}_{3}^{2}) \\
&=&\frac{1}{2m}[(p_{1}-\frac{i\hbar }{2}\frac{\partial }{\partial q_{1}})^{2}
\\
&=&+(p_{2}-\frac{i\hbar }{2}\frac{\partial }{\partial q_{2}})^{2}+(p_{3}-%
\frac{i\hbar }{2}\frac{\partial }{\partial q_{3}})^{2}]\,.
\end{eqnarray*}
respectively.

Tedious manipulations leads to the following set of
commutation relations for this set of unitary operators:

\begin{equation*}
\lbrack \widehat{L}_{i},\widehat{L}_{j}]=i\hbar \epsilon _{ijk}\widehat{L}%
_{k},
\end{equation*}%
\begin{equation*}
\lbrack \widehat{L}_{i},\widehat{K}_{j}]=i\hbar \epsilon _{ijk}\widehat{K}%
_{k},
\end{equation*}%
\begin{equation*}
\lbrack \widehat{L}_{i},\widehat{P}_{j}]=i\hbar \epsilon _{ijk}\widehat{P}%
_{k},
\end{equation*}%
\begin{equation*}
\lbrack \widehat{K}_{i},\widehat{P}_{j}]=i\hbar m\delta _{ij}\mathbf{1},
\end{equation*}%
\begin{equation*}
\lbrack \widehat{K}_{i},\widehat{H}]=i\hbar \widehat{P}_{i},
\end{equation*}%
all other commutation relations vanishing. 

This is the Galilei-Lie algebra, with a central extension
given by $m$. The operators defining the
Galilei symmetry $\widehat{P}$, $\widehat{K}$, $\widehat{L}$ and $\widehat{H}
$ are then generators of translations, boost, rotations and time
translations, respectively. To
obtain this physical content we first notice that $\widehat{Q}$ and $\widehat{P}$ transform under the boost as the physical position and
momentum, respectively, i. e.,
\begin{equation}
\exp (-i\mathbf{v}.\frac{\widehat{K}}{\hbar })\widehat{Q}_{j}\exp (i\mathbf{v%
}.\frac{\widehat{K}}{\hbar })=\widehat{Q}_{j}+v_{j}t\mathbf{1},
\label{eq 13a}
\end{equation}
\begin{equation}
\exp (-i\mathbf{v}.\frac{\widehat{K}}{\hbar })\widehat{P}_{j}\exp (i\mathbf{v%
}.\frac{\widehat{K}}{\hbar })=\widehat{P}_{j}+mv_{j}\mathbf{1}.
\label{eq 13b}
\end{equation}%

Furthermore, the operators $\widehat{Q}$ and $\widehat{P}$ does not commute
with each other, that is,
\begin{equation*}
\lbrack \widehat{Q}_{i},\widehat{P}_{j}]=i\hbar \delta _{ij}\mathbf{1}.
\end{equation*}%

$\widehat{Q}$ and $\widehat{P}$ can therefore be taken to be the position
and momentum physical observables, respectively. To
be consistent, the generators $\widehat{L}$ are interpreted as the
angular momentum observable, and $\widehat{H}$ is taken to be the
Hamiltonian operator. The Casimir invariants of the Lie
algebra are given by

\begin{equation*}
I_{1}=\widehat{H}-\frac{\widehat{P}^{2}}{2m}\quad \mathrm{and}\quad I_{2}=%
\widehat{L}-\frac{1}{m}\widehat{K}\times \widehat{P},
\end{equation*}%
where $I_{1}$ describes the Hamiltonian of a free particle and $I_{2}$ is
associated with the spin degrees of freedom. Here we are concerned with the
scalar representation; i.e. spin zero, such that $I_{2}=0$.

Defining the operators
\begin{equation*}
\overline{Q}=q\mathbf{1}\ \ \ \mathrm{and}\ \ \overline{P}=p\mathbf{1},
\end{equation*}%
we observe that, under the boost, $\overline{Q}$ and $\overline{P}$
transform as
\begin{equation*}
\exp (-iv\frac{\widehat{K}}{\hbar })2\overline{Q}\exp (iv\frac{\widehat{K}}{%
\hbar })=2\overline{Q}+vt\mathbf{1},
\end{equation*}
and
\begin{equation*}
\exp (-iv\frac{\widehat{K}}{\hbar })2\overline{P}\exp (iv\frac{\widehat{K}}{%
\hbar })=2\overline{P}+mv\mathbf{1}.
\end{equation*}%
This shows that $\overline{Q}$ and $\overline{P}$ transform as position and
momentum variables, respectively.These operators commute, i.e. $[\overline{Q},\overline{P}]=0$. Then $\overline{Q}$ and $%
\overline{P}$ and therefore cannot be interpreted
as observables. Nevertheless, they allow construction
of a Hilbert-space frame with the content of phase
space. To this end, we define an orthogonal basis in $\mathcal{H}(\Gamma )$ by
stating that $\overline{Q}|q,p\rangle =q|q,p\rangle $ and $\overline{P}%
||q,p\rangle =p|q,p\rangle $, with
\begin{equation*}
\langle q,p|q^{\prime },p^{\prime }\rangle =\delta (q-q^{\prime })\delta
(p-p^{\prime }),
\end{equation*}%
such that $\int dqdp|q,p\rangle \langle q,p|=1.$ 

Although associated with the system state, the wave
function  $\psi (q,p,t)=\langle q,p|\psi (t)\rangle $ does not have the usual
content of a quantum-mechanical state. This point deserves
a brief digression.

The time evolution of $\psi (q,p,t)$ is given by the generator
of time translations:
\begin{equation}
\psi (t)=e^{\frac{-i\widehat{H}t}{\hbar }}\psi (0),  \label{eq 12}
\end{equation}%
and its Hermitian adjoint is given by the equality
\begin{equation}
\psi ^{\dagger }(t)=\psi ^{\dagger }(0)e^{\frac{i\widehat{H}t}{\hbar }}.
\end{equation}%

We therefore obtain the result

\begin{equation*}
i\hbar \partial _{t}\psi (q,p;t)=\widehat{H}(q,p)\psi (q,p;t),
\end{equation*}%
or
\begin{equation}
i\hbar \partial _{t}\psi (q,p;t)=H(q,p)\star \psi (q,p;t),  \label{eq 13}
\end{equation}%
which is the Schr\"odinger equation in phase space~\cite{seb1}.

The expectation value of a physical observable
$\widehat{A}(q,p)=a(q,p;t)\star$, in the state  $\psi (q,p)$ is given by the
equalities

\begin{eqnarray}
\left\langle A\right\rangle &=&\int dqdp\psi ^{\dagger }(q,p)\widehat{A}%
(q,p)\ \psi (q,p)\   \notag \\
&=&\int dqdp\psi ^{\dagger }(q,p)[a(q,p)\star \psi (q,p)]  \notag \\
&=&\int dqdp\ a(q,p)[\psi (q,p)\star \psi ^{\dagger }(q,p)].  \label{bert555}
\end{eqnarray}%

To physically interpret the formalism, we associate
 $\psi (q,p,t)$ with the Wigner function, $f_{W}(q,p)$, which is
given by the expression~\cite{seb1},
\begin{equation}
f_{W}(q,p)=\psi (q,p,t)\star \psi ^{\dagger }(q,p,t).  \label{eq14}
\end{equation}%

The Wigner function satisfies the Liouville-von Neumann
equation \cite{seb1} and determines the probability density
both in configuration space,
\begin{equation}
\rho (q)=\int dp\,[\psi (q,p)\star \psi ^{\dagger }(q,p)]=\int dp\,\psi
(q,p)\psi ^{\dagger }(q,p),  \label{B21}
\end{equation}%
and in momentum space,
\begin{equation}
\rho (p)=\int dq\,[\psi (q,p)\star \psi ^{\dagger }(q,p)]=\int dq\,\psi
(q,p)\psi ^{\dagger }(q,p).  \label{b233}
\end{equation}

The expression for the expectation value of an observable
is therefore consistent with the Wigner formalism, i.e. from Eqs.~(\ref{bert555}) and (%
\ref{eq14}), we have that
\begin{equation*}
\left\langle A\right\rangle =\int dqdp\ a(q,p)f_{W}(q,p;t).
\end{equation*}%

We therefore have a complete set of physical prescriptions
to interpret the symplectic star-representations,
which paves the road to application. The following sections
discuss the three dimensional harmonic oscillator
and the non-commutative oscillator.

\section{3D Harmonic Oscillator in Phase Space}

In this section we construct the solutions of the harmonic
oscillator in phase space. Consider the following
3-dimensional Hamiltonian
\begin{equation}
H=\frac{p^{2}}{2m}+\frac{1}{2}\,m\omega ^{2}q^{2}\,,  \label{1}
\end{equation}%
where $p^{2}=p_{x}^{2}+p_{y}^{2}+p_{z}^{2}$ and $q^{2}=x^{2}+y^{2}+z^{2}$.

In phase space, we replace the coordinates and momenta
by
\begin{equation*}
q\star =q_{i}+\frac{\imath }{2}\,\frac{\partial }{\partial p^{i}}\,,\qquad
p^{i}\star =p^{i}+\frac{\imath }{2}\,\frac{\partial }{\partial q_{i}}\,,
\end{equation*}%
respectively, where we have set $m=\omega =\hbar =1$. 

Then we have that%
\begin{eqnarray*}
H\star &=&\frac{1}{2}\biggl[p_{x}^{2}+p_{y}^{2}+p_{z}^{2}+x^{2}+y^{2}+z^{2}
\\
&&+i\left( x\frac{\partial }{\partial p_{x}}-p_{x}\frac{\partial }{\partial x%
}\right) +i\left( y\frac{\partial }{\partial p_{y}}-p_{y}\frac{\partial }{%
\partial y}\right) \\
&&+i\left( z\frac{\partial }{\partial p_{z}}-p_{z}\frac{\partial }{\partial z%
}\right) \\
&&-\frac{1}{4}\,\left( \frac{\partial ^{2}}{\partial x^{2}}+\frac{\partial
^{2}}{\partial y^{2}}+\frac{\partial ^{2}}{\partial z^{2}}+\frac{\partial
^{2}}{\partial p_{x}^{2}}+\frac{\partial ^{2}}{\partial p_{y}^{2}}+\frac{%
\partial ^{2}}{\partial p_{z}^{2}}\right) \biggr]\,.
\end{eqnarray*}

To solve the equation $H\star \Psi =E\Psi $, we change variables
with the definition
\begin{equation*}
\zeta =\frac{1}{2}\left(
p_{x}^{2}+p_{y}^{2}+p_{z}^{2}+x^{2}+y^{2}+z^{2}\right) \,,
\end{equation*}%
such that
\begin{eqnarray}
\frac{\partial \Psi }{\partial x}=x\frac{\partial \Psi }{\partial \zeta }\,,
&\qquad &\frac{\partial \Psi }{\partial p_{x}}=p_{x}\frac{\partial \Psi }{%
\partial \zeta }\,,  \notag \\
\frac{\partial \Psi }{\partial y}=y\frac{\partial \Psi }{\partial \zeta }\,,
&\qquad &\frac{\partial \Psi }{\partial p_{y}}=p_{y}\frac{\partial \Psi }{%
\partial \zeta }\,,  \notag \\
\frac{\partial \Psi }{\partial z}=z\frac{\partial \Psi }{\partial \zeta }\,,
&\qquad &\frac{\partial \Psi }{\partial p_{z}}=p_{z}\frac{\partial \Psi }{%
\partial \zeta }\,,  \notag
\end{eqnarray}%
and
\begin{eqnarray}
\frac{\partial ^{2}\Psi }{\partial x^{2}}=\frac{\partial \Psi }{\partial
\zeta }+x^{2}\frac{\partial ^{2}\Psi }{\partial \zeta ^{2}}\,, &\qquad &%
\frac{\partial ^{2}\Psi }{\partial p_{x}^{2}}=\frac{\partial \Psi }{\partial
\zeta }+p_{x}^{2}\frac{\partial \Psi }{\partial \zeta }\,,  \notag \\
\frac{\partial ^{2}\Psi }{\partial y^{2}}=\frac{\partial \Psi }{\partial
\zeta }+y^{2}\frac{\partial ^{2}\Psi }{\partial \zeta ^{2}}\,, &\qquad &%
\frac{\partial ^{2}\Psi }{\partial p_{y}^{2}}=\frac{\partial \Psi }{\partial
\zeta }+p_{y}^{2}\frac{\partial ^{2}\Psi }{\partial \zeta ^{2}}\,,  \notag \\
\frac{\partial ^{2}\Psi }{\partial z^{2}}=\frac{\partial \Psi }{\partial
\zeta }+z^{2}\frac{\partial ^{2}\Psi }{\partial \zeta ^{2}}\,, &\qquad &%
\frac{\partial ^{2}\Psi }{\partial p_{z}^{2}}=\frac{\partial \Psi }{\partial
\zeta }+p_{z}^{2}\frac{\partial ^{2}\Psi }{\partial \zeta ^{2}}\,.  \notag
\end{eqnarray}%

Consequently, the imaginary part of $H\star \Psi =E\Psi $ vanishes.
This leads to
\begin{equation}
\zeta\frac{\partial^2\Psi}{\partial \zeta^2}+3\frac{\partial\Psi}{\partial
\zeta}-4(\zeta-E)\Psi=0\,.
\end{equation}

To change variables again, we define $r=4\zeta $, write
$\Psi=\exp {\left( -r/2\right) }\,\chi (r)$, and obtain the following equation:

\begin{equation}
r\chi ^{\prime \prime }+(3-r)\chi ^{\prime }+(E-\frac{3}{2})\chi =0\,,
\label{3}
\end{equation}%
where the prime indicates differentiation with respect to $r$. 

Equation (\ref{3}) is of the general form

\begin{equation*}
zu^{\prime \prime }+(\gamma -z)u^{\prime }-\alpha u=0\,,
\end{equation*}%
where $u=u(z)$, a solution of which is the confluent hypergeometric
function, defined by the equality
\begin{eqnarray*}
u(z) &=&F(\alpha .\gamma ,z)=1+\frac{\alpha }{\gamma }z+\frac{\alpha (\alpha
+1)}{\gamma (\gamma +1)}\frac{z^{2}}{2} \\
&&+\frac{\alpha (\alpha +1)(\alpha +2)}{\gamma (\gamma +1)(\gamma +2)}\frac{%
z^{3}}{6}+\dots .
\end{eqnarray*}

Comparison with Eq.~(\ref{3}) now shows that $\chi
=F\left( -\left( E-\frac{3}{2}\right) ;3;r\right) $. The confluent hypergeometric function
is finite if the parameter $\alpha $ is a negative integer. This
constraint yields the result
\begin{equation*}
E=E_{n}=n+\frac{3}{2}\,,
\end{equation*}%
where $n$ is an integer, which is the expected result for the
energy, since $\hbar =\omega=1$. 

The solution of the phase-space Schr\"{o}dinger equation is

\begin{equation}
\Psi _{n}(\zeta )=\exp {(-2\zeta )}\,F(-n,3,4\zeta )\,,
\end{equation}%
where $F(-n,3,4\zeta )$ is the con
uent hypergeometric function
with the appropriate parameters.

The Wigner function is calculated from Eq. (\ref{eq14}), which
in this case reads $f_{W}^{n}(q,p)=\Psi _{n}(\zeta )\star \Psi _{n}(\zeta )$. We therefore
have that
\begin{equation}
f_{W}^n(q,p)=C_n\,\exp{(-2\zeta)}\,F(-n,3,4\zeta)\,,
\end{equation}
where $C_n=\exp{(-2E_n)}\,F(-n,3,4E_n)$.

\section{2D Non-commutative Oscillator in Phase Space}

We now want to derive the Wigner function for a two-dimensional
non-commutative oscillator in phase space,
which is defined by the Hamiltonian
\begin{equation}
H=\frac{1}{2}(x^{2}+p_{x}^{2})+\frac{1}{2}(y^{2}+p_{y}^{2})\,,  \label{2d1}
\end{equation}%
where again $\hbar =1$, $m=1$ and $\omega =1$.

The star-product is now given by the expression
\begin{eqnarray}
\star &=&\star _{\hbar \theta }=\mathrm{exp}\left\{ \frac{i\ }{2}%
\sum_{i=1}^{2}(\overleftarrow{\partial }_{q_{i}}\overrightarrow{\partial }%
_{p_{i}}-\overleftarrow{\partial }_{p_{i}}\overrightarrow{\partial }%
_{q_{i}})\right.  \notag \\
&+&\left. \frac{i\theta }{2}(\overleftarrow{\partial }_{x}\overrightarrow{%
\partial }_{y}-\overleftarrow{\partial }_{y}\overrightarrow{\partial }%
_{x})\right\}    \notag
\end{eqnarray}%
where $q_{i}=(x,y)$ and $p_{i}=(p_{x},p_{y})$. 

The position and momentum operators are given by
the expressions
\begin{equation}
q_{i}\star =q_{i}+\frac{i}{2}\partial _{p_{i}}+\frac{i}{2}\theta
_{ij}\partial _{q_{j}}\,, \label{2d2}
\end{equation}%
and
\begin{equation}
p_{i}\star =p_{i}+\frac{i}{2}\partial _{q_{i}}+\frac{i}{2}\theta
_{ij}\partial _{p_{j}}\,,  \label{2d3}
\end{equation}%
respectively.

These operators satisfy the following nonzero commutation
relations
\begin{equation*}
\lbrack q_{i},p_{j}]=i\delta _{ij},[q_{i},q_{j}]=i\theta
_{ij},[p_{i},p_{j}]=-i\theta _{ij}.
\end{equation*}%

In addition to the usual phase-space non-commutativity,
we therefore have momentum coordinates
that do not commute and space coordinates
that do not commute. From Eqs. (\ref{2d2}) and
Eq. (\ref{2d3}), we obtain the following Schr\"{o}dinger equation,
$H\star \psi (x,y,p_{x},p_{y})=E\psi (x,y,p_{x},p_{y})$:
\begin{eqnarray*}
E\psi (x,y,p_{x},p_{y}) &=&\frac{1}{2}[(x+\frac{i}{2}\partial _{p_{x}}+\frac{%
i}{2}\theta \partial _{y})^{2} \\
&&+(p_{x}+\frac{i}{2}\partial _{x}-\frac{i}{2}\theta \partial _{p_{y}})^{2}
\\
&&+(y+\frac{i}{2}\partial _{p_{y}}-\frac{i}{2}\theta \partial _{x})^{2} \\
&&+(p_{y}+\frac{i}{2}\partial _{y}+\frac{i}{2}\theta \partial
_{p_{x}})^{2}]\psi (x,y,p_{x},p_{y}),
\end{eqnarray*}%
where we have used that $\theta =\theta _{12}=-\theta _{21}$.

To solve this Schr\"odinger equation, we define the coordinates
$\widetilde{x}=x$%
, $\widetilde{y}=(1+\theta ^{2})^{-1/2}(y\star -\theta p_{x})$, $\widetilde{%
p_{x}}\star =(1+\theta ^{2})^{-1/2}(p_{x}\star +\theta y\star )$, $%
\widetilde{p_{y}}\star =p_{y}\star $ and the star-operators

\begin{equation*}
\widetilde{x}\star=x\star,
\end{equation*}
\begin{equation*}
\widetilde{y}\star=(1+\theta^{2})^{-1/2}(y\star-\theta p_{x}\star),
\end{equation*}
\begin{equation*}
\widetilde{p_{x}}\star =(1+\theta ^{2})^{-1/2}(p_{x}\star +\theta y\star ),
\end{equation*}%
and
\begin{equation*}
\widetilde{p_{y}}\star =p_{y}\star .
\end{equation*}
The latter satisfy the following commutation relations
\begin{equation*}
\lbrack \widetilde{x}\star ,\widetilde{p_{x}}\star ]=(1+\theta ^{2})^{1/2},[%
\widetilde{y}\star ,\widetilde{p_{y}}\star ]=(1+\theta ^{2})^{1/2}.
\end{equation*}
We therefore define the annihilation operators
\begin{eqnarray*}
\widetilde{a}_{x}\star &=&\frac{1}{\sqrt{2}}(\widetilde{x}\star +i\widetilde{%
p_{x}}\star )\,, \\
\widetilde{a}_{y}\star &=&\frac{1}{\sqrt{2}}(\widetilde{y}\star +i\widetilde{%
p_{y}}\star )\,,
\end{eqnarray*}
and the creation operators
\begin{eqnarray*}
\widetilde{a}_{x}^{\dagger }\star &=&\frac{1}{\sqrt{2}}(\widetilde{x}\star -i%
\widetilde{p_{x}}\star )\,, \\
\widetilde{a}_{y}^{\dagger }\star &=&\frac{1}{\sqrt{2}}(\widetilde{y}\star +i%
\widetilde{p_{y}}\star )\,,
\end{eqnarray*}
such that $[\widetilde{a}_{i}\star ,\widetilde{a}_{j}^{\dagger }\star
]=i(1+\theta ^{2})^{1/2}\delta _{ij}$, where $\widetilde{a}_{1}\star =%
\widetilde{a}_{x}\star $ and $\widetilde{a}_{2}\star =\widetilde{a}_{y}\star
$. 

The Schr\"{o}dinger equation can then be written in the
form
\begin{eqnarray*}
H\star \psi (x,y,p_{x},p_{y}) &=&E\psi (\widetilde{x},\widetilde{p_{x}},%
\widetilde{y},\widetilde{p_{y}}) \\
&=&[\widetilde{a}_{x}^{\dagger }\star \widetilde{a}_{x}\star +\widetilde{a}%
_{y}^{\dagger }\star \widetilde{a}_{y}\star \\
&&+(1+\theta ^{2})^{1/2}]\psi (\widetilde{x},\widetilde{p_{x}},\widetilde{y},%
\widetilde{p_{y}}).
\end{eqnarray*}%

The energy eigenvalues are then given by the expression
\begin{equation*}
E_{n_{x}n_{y}}=(1+\theta ^{2})^{1/2}(n_{x}+n_{y}+1)\,.
\end{equation*}

For the ground state,  $\psi _{00}(\widetilde{x},\widetilde{p_{x}},\widetilde{y},%
\widetilde{p_{y}})=\phi _{0}(\widetilde{x},\widetilde{p_{x}})\chi _{0}(%
\widetilde{y},\widetilde{p_{y}})$, and we have the equations
$\widetilde{a}%
_{x}\star \phi _{0}=\widetilde{a}_{y}\star \chi _{0}=0$, which can be explicitly
written as

\begin{equation}
\frac{1}{\sqrt{2}}(\widetilde{x}+\frac{i}{2}\partial _{\widetilde{p_{x}}}+i%
\widetilde{p_{x}}+\frac{1}{2}\partial _{\widetilde{x}})\phi (\widetilde{x},%
\widetilde{p_{x}})=0,  \label{2d6}
\end{equation}%
and
\begin{equation}
\frac{1}{\sqrt{2}}(\widetilde{y}+\frac{i}{2}\partial _{\widetilde{p_{y}}}+i%
\widetilde{p_{y}}+\frac{1}{2}\partial _{\widetilde{y}})\phi (\widetilde{y},%
\widetilde{p_{y}})=0.  \label{2d7}
\end{equation}%
To find real solutions, we have to solve the following
set of equations:
\begin{eqnarray*}
(\widetilde{x}+\frac{1}{2}\partial _{\widetilde{x}})\phi _{0} &=&0, \\
(\widetilde{y}+\frac{1}{2}\partial _{\widetilde{y}})\chi _{0} &=&0, \\
(\widetilde{p_{x}}+\frac{1}{2}\partial _{\widetilde{p_{x}}})\phi _{0} &=&0,
\\
(\widetilde{p_{y}}+\frac{1}{2}\partial _{\widetilde{p_{y}}})\chi _{0} &=&0.
\end{eqnarray*}

The general ground-state solution is given by the expression
\begin{equation}
\psi _{00}=C_{0}\exp {-(\widetilde{x}^{2}+\widetilde{p_{x}}^{2}+\widetilde{y}%
^{2}+\widetilde{p_{y}}^{2})},  \label{2d8}
\end{equation}%
where $C_{0}=\frac{1}{\pi }$ is the normalization constant. 

For $n\geq 1$ the functions $\psi _{n}$ are determined by the
creation operator, that is, by the relation
\begin{equation}
\psi _{n_{x}n_{y}}=\frac{1}{\sqrt{n!}}(\widetilde{a}_{x}^{\dagger }\star
\widetilde{a}_{y}^{\dagger }\star )^{n}\psi _{00}.  \label{2d9}
\end{equation}%

The Wigner function associated with each $\psi _{n_{x}n_{y}}$ is
\begin{equation*}
f_{W}(\widetilde{x},\widetilde{p_{x}},\widetilde{y},\widetilde{p_{y}})=\psi
_{n_{x}n_{y}}\star \psi _{n_{x}n_{y}}^{\dagger }.
\end{equation*}%

In particular, for $n_{x}=1,n_{y}=1$, we find that
\begin{eqnarray*}
f_{W}^{1}(\widetilde{x},\widetilde{p_{x}},\widetilde{y},\widetilde{p_{y}})\
&\sim &[1-2((\widetilde{x})^{2}+(\widetilde{p_{x}})^{2})]e^{-((\widetilde{x}%
)^{2}+(\widetilde{p_{x}})^{2})} \\
&&\times \lbrack 1-2((\widetilde{y})^{2}+(\widetilde{p_{y}})^{2})]e^{-((%
\widetilde{y})^{2}+(\widetilde{p_{y}})^{2})}\,,
\end{eqnarray*}%
and for $n_{x}=2,n_{y}=2$,
\begin{eqnarray*}
f_{W}^{2}(\widetilde{x},\widetilde{p_{x}},\widetilde{y},\widetilde{p_{y}})
&\sim &[2-4((\widetilde{x})^{2}+(\widetilde{p_{x}})^{2})+((\widetilde{x}%
)^{2}+(\widetilde{p_{x}})^{2})^{2}] \\
&&\times e^{-((\widetilde{x})^{2}+(\widetilde{p_{x}})^{2})}[2-4((\widetilde{y%
})^{2}+(\widetilde{p_{y}})^{2}) \\
&&+((\widetilde{y})^{2}+(\widetilde{p_{y}})^{2})]e^{-((\widetilde{y})^{2}+(%
\widetilde{p_{y}})^{2})}.
\end{eqnarray*}
{\small \ } 

For arbitrary $n_{x}$ and $n_{y}$ we have the result
\begin{eqnarray*}
f_{W}^{n} &\sim &Ln[(\widetilde{p_{x}})^{2}+(\widetilde{x})^{2})] \\
&&\times Ln[(\widetilde{y})^{2}+(\widetilde{p_{y}})^{2}]e^{-((\widetilde{x}%
)^{2}+(\widetilde{p_{x}})^{2}+(\widetilde{y})^{2}+(\widetilde{p_{y}})^{2})},
\end{eqnarray*}%
where $Ln$ are the Laguerre polynomials. 

Going back to the original variables, we have that
\begin{eqnarray*}
f_{W}^{n}(x,y,p_{x},p_{y};\theta ) &\sim &Ln[(x^{2}+(1+\theta
^{2})^{-1}(p_{x}+\theta y)^{2})] \\
&\times &Ln[(1+\theta ^{2})^{-1}(y-\theta p_{x})^{2}+p_{y}^{2}] \\
&\times &e^{-((x^{2}+\beta (p_{x}+\theta y)^{2})+\beta (y-\theta
p_{x})^{2}+p_{y}^{2})}\,,\label{27}
\end{eqnarray*}%
where $\beta =(1+\theta ^{2})^{-1}$. 

Equation (\ref{27}) was derived, by a different method, in
Refs. ~\cite{ncw} and \cite{ncw2}. It is important to notice that additional
solutions can be found, associated with different
combinations of the quasi-amplitudes of probability. The
Wigner function for the non-commutative oscillator depends
on the parameter $\theta $, which finds physical application
in such problems as the quantum Hall effect~\cite{teta1,teta2,teta3}.

\section{Concluding remarks}

We have set forth a symplectic representation of the
Galilei group, which yields quantum theories in phase
space. We have derived a Schr\"{o}dinger equation and, as
illustrations, studied the 3D harmonic oscillator and the
non-commutative oscillator in phase space. In both cases,
we obtained the Wigner functions. The symplectic representation
is constructed on the basis of the Moyal- or
star-product, an ingredient of non-commutative geometry.
A Hilbert space is then defined from a manifold with
the features of phase space. The states are represented
by a quasi-amplitude of probability, a wave function in
phase space, the definition of which makes connection
with the Wigner function, i. e., the quasi-probability density.
Non-trivial, yet consistent, the association with the
Wigner function provides a physical interpretation of the
theory. Analogous interpretations are not found in other
studies of representations in phase space~\cite{seb42,seb43}.

One aspect of the procedure deserves emphasis. Our
formalism explores unitary representations to calculate
Wigner functions. This constitutes an important advantage
over the more traditional constructions of the
Wigner method, which entail several intricacies associated
with the Liouville-von Neumann equation. Furthermore,
the formalism we have described opens new perspectives
for applications of the Wigner-function method
in quantum field theory. This aspect of the formalism
will be discussed in a forthcoming paper.

\textbf{Acknowledgements: }This work was partially supported by CAPES and
CNPq.

\end{document}